\documentstyle[prl,epsf,aps]{revtex}

\newcommand \be {\begin{equation}}
\newcommand \bea {\begin{eqnarray}}
\newcommand \ee {\end{equation}}
\newcommand \eea {\end{eqnarray}}

\makeatletter
\newcommand\erfc{\mathop{\operator@font erfc}\nolimits}
\makeatother
\input{epsf}
\begin{document}
\twocolumn[\hsize\textwidth\columnwidth\hsize\csname@twocolumnfalse\endcsname
\draft

\title{Exactly solvable phase oscillator models with synchronization
dynamics}

\author{L. L. Bonilla} \address{Escuela Polit{\'e}cnica Superior,
Universidad Carlos III de Madrid,
Butarque 15, \\ 28911 Legan{\'e}s, Spain. E-Mail: bonilla@ing.uc3m.es}

\author{C. J. P{\'e}rez Vicente and F. Ritort} \address{Departament de
Fisica Fonamental, Facultat de Fisica, Universitat de Barcelona,
Diagonal 647\\ 08021 Barcelona, Spain. E-Mail:
conrad@ffn.ub.es,ritort@ffn.ub.es} \author{J. Soler}
\address{Departamento de Matem{\'a}tica Aplicada, Facultad de Ciencias,
Universidad de Granada \\ 18071 Granada, Spain. E-mail: jsoler@ugr.es}

\date{\today}
\maketitle

\begin{abstract}
Populations of phase oscillators interacting globally through a general
coupling function $f(x)$ have been considered. In the absence of precessing
frequencies and for odd-coupling functions there exists a Lyapunov
functional and the probability density evolves toward stable
stationary states described by an equilibrium measure. We have
then proposed a family of exactly solvable models with singular
couplings which synchronize more easily as the coupling
becomes less singular. The stationary solutions of the least singular
coupling considered, $f(x)=$ sign$(x)$, have been found analytically in terms
of elliptic functions. This
last case is one of the few non trivial models for synchronization
dynamics which can be analytically solved.
\end{abstract}

\vfill

\vfill

\twocolumn
\vskip.5pc]

\narrowtext
The dynamics of systems consisting of large sets of mutual interacting
units is a very interesting problem in statistical physics. In some cases,
these units can be thought of as subsystems characterized by hidden internal
degrees of freedom obeying their own dynamics. This is the case of large
populations of individuals forming complex systems of interest in
interdisciplinary fields such as biology, economy, neurophysiology and ecology
\cite{winfree,Stro94}.
On the other hand, the subsystems can be seen as entities interacting with each
other in the presence of quenched disordered external fields. The effect
of the external forces is to pump energy into the system thereby leading to
nonlinear oscillatory behavior. This last case describes charged wave 
instabilities in plasmas \cite{PL}, dynamical response of Josephson
junction arrays in the presence of an external ac field \cite{wiesen} or
nonlinear oscillations and coherent motion of magnetized domains in
strongly coupled magnetic systems \cite{RI}. It was realized by Kuramoto
\cite{kuramoto} that one could realize this rich dynamical behavior by
considering mean-field models of phase oscillators with randomly quenched
natural frequencies. The crucial feature of the Kuramoto model is that the
random precession frequencies play the role of external forces which pump
energy
into the system. This leads to dynamical instabilities and synchronization in
low dissipation regimes. Moreover, the mean-field character of the model
captures the essential mechanism which leads to synchronized dynamics while
retaining mathematical simplicity.

In the simplest setting, we consider the dynamics of a system of
nonlinearly globally coupled phase oscillators with random frequencies
$\omega_i$ taken from a distribution $g(\omega)$ and subject to external
independent white noise sources $\eta_{i}$ (of strength $\sqrt{2T}$):
\begin{eqnarray}
\frac{\partial\phi_i}{\partial t}= \omega_{i} -
\frac{K_0}{N}\sum_{j=1}^N f(\phi_i-\phi_j) + \eta_i(t),
\label{eq1}
\end{eqnarray}
$i=1,\ldots,N$. Here $\phi_{i}(t)$ denotes the {\it i}th oscillator phase,
$K>0$ represents the coupling strength and $f$ is a generic real function
of periodicity $2\pi$. In Fourier space, the latter can be decomposed as
$f(x)=\sum_{n=-\infty}^{\infty} a_n e^{inx} ,$
with $ a_n=a_{-n}^*$.

Important work on the Kuramoto model with a generalized coupling
function was carried out by Daido \cite{daido} who introduced the concept of
order function and Crawford \cite{Cr} who proposed scaling behavior of
bifurcating branches from the incoherent solution. These works introduced
rather general theories and methods which, however, had inherent
limitations (zero temperature for the order function theory and vicinity to
bifurcation points in Crawford's work). Thus it would be desirable
to have exact results for solvable models where such theories could be checked
and extended.

The purpose of this letter is to show some exact results on the
synchronization dynamics of simple models of oscillators.  Exact results
are very difficult to obtain in models with general smooth coupling
functions $f(x)$. Here we will follow a different strategy by analyzing
models with singular couplings which can be mapped into well known
physical problems. Later we will go further and consider models with
smoother $f(x)$ such as the Daido coupling which we analytically
solve. While our analysis is restricted to models without disorder, the
techniques developed here could be extended to disordered cases \cite{acebron}.

{\it The moment approach and general equations.}
A simple way to study the dynamics of the Kuramoto model
with general coupling functions (\ref{eq1}) is to use the recently
introduced moment approach \cite{PR} and define,
$H_{k}^m=\frac{1}{N}\sum_{j=1}^{N}\,\overline{<e^{ik\phi_{j}}>\omega_j^m}\,,
$
where the brackets $<...>$ denote average with respect to the external noise
and the overbar denotes average with respect to the random oscillator frequency.
The equations of motion for the moments read
\begin{eqnarray}
\frac{\partial H_{k}^m}{\partial t}=-K_0 i k
\sum_{n=-\infty}^{\infty} a_n H_{k+n}^m H_{-n}^0 \nonumber\\
- k^2 T H_{k}^m\, + \,ik H_{k}^{m+1} .
\label{eq4}
\end{eqnarray}
The following differential equation for the generating function,
$g(x,y,t)= \sum_{k=-\infty}^{\infty}\sum_{m=0}^{\infty}
e^{-ikx}\frac{y^m}{ 2\pi\, m!}H_k^m(t) ,
$ is easy to derive from (\ref{eq4})
\begin{eqnarray}
\frac{\partial g}{\partial t} = -\frac{\partial}{\partial x}
\Bigl [ v(x,t)\, g \Bigr ] + T\frac{\partial^2 g}{\partial
x^2}-\frac{\partial^2 g }{\partial x\partial y} \,.
\label{eq6}
\end{eqnarray}
Here $v(x,t)$ is the drift velocity, defined by
\begin{eqnarray}
v(x,t) &=& -K_0 \sum_{n=-\infty}^{\infty}a_n H_{-n}^0 e^{inx} \nonumber\\
&=& -K_0 \int_{-\pi}^{\pi} f(x') g(x-x',0,t) \, dx' .
\label{eq8}
\end{eqnarray}
It is a simple matter to check that this velocity satisfies $v(x,t) = - K_0 \,
H(x-\Omega_e t,t)$, where $H(x,t)$ is Daido's order function~\cite{daido}
and $\Omega_e$ is the synchronization frequency (which is
zero for stationary states). On the other hand, the generating function
$g(x,y,t)$ is related to the one-oscillator probability density
$\rho(x,\omega,t)$ through
$g(x,y,t)=\int_{-\infty}^{\infty} e^{y\omega}\,\rho(x,\omega,t)\, g(\omega)\,
d\omega.$
We derive from (\ref{eq6}) the usual nonlinear Fokker-Planck equation for
$\rho(x,\omega,t)$ wherever $g(\omega)\ne 0$
\begin{eqnarray}
\frac{\partial \rho}{\partial t} + \frac{\partial}{\partial x}
\Bigl [ (\omega + v)\, \rho \Bigr ] = T\frac{\partial^2 \rho}{\partial
x^2} \,. \label{eq10a}
\end{eqnarray}

{\em Existence of a Lyapunov functional.}
The solution of (\ref{eq10a}) is generally quite complicated: only in
special cases it is possible to work out explicit results. Here we
want to analyze under which conditions it is possible to find
a Lyapunov functional for the dynamics. This is an important result since
such Lyapunov functional explicitly yields a functional equation for the
stationary states as well as stability results. In particular we will find
a Lyapunov functional when the coupling $f(x)$ is an odd function (i.e.\ if
detailed balance is satisfied) and there is no frequency disorder [i.e.\
$g(\omega)=\delta(\omega)$]. Then there exists a partition function which
characterizes the thermodynamic properties of the stationary states.

Let us define the potential function
$V(x,t) = \int_{-\pi}^x v(s,t)\, ds$
and restrict ourselves to finding stationary solutions; rotating
wave solutions may be reduced to this case after moving to a rotating
frame. It is possible to show that the stationary solutions should have the
form
\begin{eqnarray}
\rho(x,\omega) & = & Z^{-1}\, e^{{(\pi + x)\, \omega + V(x)\over T}}
\nonumber\\
& - &{J\over T} \int_{-\pi}^x
\exp\left[ {(x-s)\omega + V(x) - V(s)\over T}\right]\, ds ,\label{eq12}
\end{eqnarray}
where $Z$ and $J$ are independent of $x$. We now impose the condition that
$\rho$ be a $2\pi$-periodic function of $x$, use the symmetry properties of
the drift and find the probability flux $J$ as a function of $Z$:
$J/T = 2Z^{-1} \mbox{sinh}(2\pi T^{-1}\omega)/\int_{-\pi}^{\pi}
e^{-[\pi +x)\omega + V(x)]/T}\, dx.$
$Z$ can be found from the normalization condition $\int_{-\pi}^{\pi}
\rho dx =1$. An interesting case corresponds to the case without
disorder, $g(\omega) = \delta(\omega)$, for which $J=0$, $\rho(x,0,t) =
g(x,y,t)\equiv g(x,t)$. We can obtain the stationary solutions (subscript
zero below) by solving the system of equations:
\begin{eqnarray}
g_0(x) = Z^{-1} e^{{V_{0}(x)\over T}}\,,\quad Z = \int_{-\pi}^{\pi}
e^{{V_{0}(s)\over T}}\, ds\,, \label{eq14}\\
v_0(x) = - K_0 \int_{-\pi}^{\pi} f(x-x') g_0(x') dx' .\label{eq141}
\end{eqnarray}

It is easy to check that, for this type of potential solutions, there is
a Lyapunov functional~\cite{Htheorem} defined by the relative entropy,
\begin{eqnarray}
\eta(t) = \int_{-\pi}^{\pi} g(x,t)\,\ln\left({g(x,t)\over \overline{g}(x,t)}
\right)\, dx ,\label{odd6}\\
\overline{g}(x,t) = e^{{V(x,t) - \mu(t)\over T}}\,,\label{eq15}\\
{d\mu\over dt} = \int_{-\pi}^{\pi} g(x,t)\, {\partial V(x,t)\over \partial t}
\,  dx .\label{eq16}
\end{eqnarray}
Direct calculation (to be presented elsewhere) shows that $\eta(t)$ is 
bounded from below, $\eta'(t)\leq
0$ and that $g$ tends to a stationary solution which is proportional to
$\overline{g}$.
We then
conclude that, for odd coupling functions and in absence of disordered
frequencies, the stationary state (\ref{eq14}) corresponds to the
thermodynamic equilibrium state of a system of $N$ oscillators interacting
through the Hamiltonian
\begin{eqnarray}
{\cal H}=\frac{K_0}{N}\,\sum_{i<j}\varepsilon(\phi_i-\phi_j) ,
\label{eq18}
\end{eqnarray}
where $\varepsilon(x)$ is a pair interaction energy $2\pi$-periodic function
defined by $\varepsilon(x)=\int_{-\pi}^x f(s) ds$. Note that
$V(x)=-K_0\int_{-\pi}^{\pi} g(x) \varepsilon(x) dx \ +$ constant. Thus a
computation of the
partition function for (\ref{eq18}) yields the stationary states.
It is important to stress that potential solutions of the type (\ref{eq14})
are no longer stationary solutions of the dynamics in the presence of
disordered frequencies $\omega_i$ \cite{PR} or when detailed balance is violated
(i.e. $f(x)$ is not an odd function).  The physical meaning of
these two conditions is quite clear: suppose a single oscillator performes a
global rotation of angle $2\pi$. The global energy of the system, given by
(\ref{eq18}), does not change because $\varepsilon(x)$ is $2\pi$-periodic. Thus
the coupling strength does not exert work into the system of oscillators when
a rotation path ($\phi_i\to\phi_i+2\pi$) is performed. Indeed, the amount
of work is $W=\int_{\phi_i}^{\phi_i+2\pi} [\omega_{i} - \frac{K_0}{N}
\sum_{j=1}^N f(x-\phi_j)] dx$ which vanishes only if $\omega_i=0$
and $f(x)=-f(-x)$.

{\em Singular coupling models.}
For general coupling functions, with infinitely many non-vanishing Fourier
modes $a_n$, it is difficult to find explicit analytical expressions for the
stationary states. But calculations turn out to be much simpler in the
case of singular coupling functions. Here we analyze three different
cases of singular interactions: a) $f(x)=\delta'(x)$; b)
$f(x)=\delta(x)$ and c) $f(x)=$ sign$(x)$. In the first case we have
$v(x)=-K\, \partial g/\partial x$,
\begin{eqnarray}
\frac{\partial g}{\partial t} = K_0\frac{\partial}{\partial x}
\Bigl (g\frac{\partial g}{\partial x}\Bigr ) + T \frac{\partial^2
g}{\partial x^2}\,.\label{eq20}
\end{eqnarray}
This equation is related to porous media systems with the additional
$2\pi$-periodicity condition for $g$. It is easy to check that
$\eta(t)=(1/2)\int_{-\pi}^{\pi} g^2(x,t) dx$ is a Lyapunov
functional. In fact, it is bounded from below and satisfies,
$\eta'(t) = - \int_{-\pi}^{\pi} (T + K_0 g)\, (\partial g/\partial x)^2 dx
\leq 0. $
Then the incoherent solution $g(x)=1/2\pi$ is globally stable and model a)
does not syncronize at any $T$.

Case b) is described by the Burgers equation
\begin{eqnarray}
\frac{\partial g}{\partial t} = 2 K_0 g\frac{\partial g}{\partial x}
+ T \frac{\partial^2 g}{\partial x^2}\,,\label{eq21b}
\end{eqnarray}
for a $2\pi$-periodic $g$, and it can be solved by the Hopf-Cole
transformation.
$\eta(t)$ defined for case a) is also a Lyapunov functional for case b), but
now
$\eta'(t) = - T \int_{-\pi}^{\pi} (\partial g/\partial x)^2 dx
\leq 0.$
At finite $T$ incoherence is again stable. But at zero temperature 
the oscillators remain synchronized forever if they are initially synchronized.

{\em Analytical solution of the model with Daido coupling}. Model c) is analyzed
carefully. This model was originally proposed by Daido in the
general disordered case but we will solve it for $g(\omega)=\delta(\omega)$.
Since $f(x)$ is odd, it is in principle possible to find the stationary states
by solving the functional equations (\ref{eq14})-(\ref{eq141}). Such a
calculation is quite involved and here we follow a different and novel
approach.
The dynamical equations for model c) are non-local because the drift velocity
(\ref{eq8}) is
\begin{eqnarray}
v(x,t) = -  K_0 \int_{0}^{\pi} [g(x-\xi,t) - g(x- \xi+\pi,t)] d\xi, \label{eq22}
\end{eqnarray}
but they become local after inserting the definitions
\begin{eqnarray}
\rho(x,t) = 2K_{0}\, [g(x,t) - g(x+\pi,t)],\nonumber\\
\sigma(x,t) = 2K_{0}\, [g(x,t) + g(x+\pi,t)] .\label{eq25}
\end{eqnarray}
The new system is
\begin{eqnarray}
\frac{\partial \rho}{\partial t} = - \frac{\partial (\sigma v)}{\partial
x} + T \frac{\partial^2 \rho}{\partial x^2}\,,\label{rho}\label{eq23}\\
\frac{\partial \sigma}{\partial t} = \frac{\partial}{\partial x} \Bigl
(v\frac{\partial v}{\partial x}\Bigr ) + T \frac{\partial^2
\sigma}{\partial x^2}\,.\label{eq24}
\end{eqnarray}

The drift velocity $v(x,t)$ is given by $\partial v/\partial x =
-\rho$. All functions are $2\pi$-periodic and it is easy to check
that both $\rho$ and $v$ are antiperiodic if $x\to x+\pi$, while $\sigma$
is $\pi$-periodic. The stationary solutions of (\ref{eq23})-(\ref{eq24})
satisfy the following equations
\begin{eqnarray}
T\sigma + {v^{2}\over 2} = L\label{eq26a}\\
T^2 \, \frac{d^2 v}{dx^2} + \left( L -
{v^{2}\over 2}\right)\, v = 0, \label{eq26b}
\end{eqnarray}
where $L$ is a positive constant. (\ref{eq26b}) may be
integrated once yielding
\begin{eqnarray}
{1\over 2}\, \left(\frac{dw}{d\xi}\right)^2 +
{w^{2}\over 2} - {w^{4}\over 8} =  {\cal E},\label{eq27}
\end{eqnarray}
where $w(\xi) =v(x)/\sqrt{L}\,,x = T\xi/\sqrt{L}$ and ${\cal E}$ is a
new constant. (\ref{eq27}) can be easily solved by quadratures in terms of
Jacobi elliptic functions \cite{abramowitz}. The periodicity properties of
the drift $v$ yields finitely many solutions (characterized by their value of
${\cal E}$ and an odd integer $n$) for a fixed value of $K_0/T$. The relation
between ${\cal E}$, $n$ and $K_0/T$ is:
\begin{eqnarray}
{4 n^{2}\, K(m)\over\pi}\, \left[ E(m)
 - {\sqrt{1-2{\cal E}}\, K(m)\over 1+\sqrt{1-2{\cal E}}}\right]
= {K_{0}\over T}\, . \label{eq28}
\end{eqnarray}
Here $K(m)$ and $E(m)$ are the complete elliptic integrals of the first and
second kind with parameter $m=(1-\sqrt{1-2{\cal E}})/(1+\sqrt{1-2{\cal E}})$,
respectively. For every admissible value of 
${\cal E}$ and $n$, with $K_0/T$ fixed, the drift velocity and the
probability density can be found from
(\ref{eq25}), (\ref{eq26a}), (\ref{eq26b}), (\ref{eq27})  and the condition
$\int_0^\pi
\sigma(x) dx = 2 K_0$. They are
\begin{eqnarray}
v(x) &=& {4nT\sqrt{m} K(m)\over\pi}\, \mbox{sn}u\,,\label{eq30}\\
g(x) &=& {K(m)\over 2\pi}\, { {1\over 1+\sqrt{1-2{\cal E}}} -
m\, \mbox{sn}^2 u - \sqrt{m}\,\mbox{cn} u\,\mbox{dn} u\over E(m)
- {\sqrt{1-2{\cal E}}\, K(m)\over 1+\sqrt{1-2{\cal E}}}}\,.\label{eq31}
\end{eqnarray}
Here $u = 2n K(m)\, (x-x_0)/\pi$, $x_0=$ constant, while sn, cn and dn are
Jacobi elliptic functions defined in \cite{abramowitz}. The details of these
calculations will be presented elsewhere.

As $K_0/T\to\infty$ the number of possible stationary solutions becomes
infinite. Each different stationary state belongs to a different solution
branch which bifurcates from incoherence at a critical value of the coupling
$K_0=n^2T\pi/2$, where $n=2p-1$ and $p\ge 1$ is an integer, see figure
1. The first branch
bifurcates from incoherence at $K_0/T=\pi/2$ and remains stable for all larger
$K_0/T$. In the limit $K_0/T\to\infty$, $v(x)/K_0 = $ sign$(x-x_0)$, $g(x)=
\delta(x-x_0-\pi)$, and full synchronization is achieved. A convenient
synchronization order parameter $r$ is defined through the global magnetization
$M=r e^{i\alpha}=(1/N)\sum_{j=1}^N e^{i\phi_{j}}$. 

\begin{figure}
\begin{center}
\leavevmode
\epsfysize=200pt{\epsffile{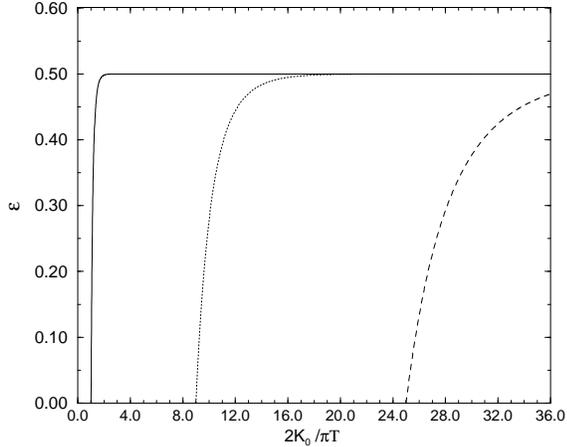}}
\end{center}
  \protect\caption[2]{Bifurcation diagram ${\cal E}$ versus $2K_{0}/(\pi
T)$ for stationary synchronized solutions branching off from incoherence
${\cal E} = 0$ at the square of the odd integer numbers.
\protect\label{FIG1} }
\end{figure}

\begin{figure}
\begin{center}
\leavevmode
\epsfysize=240pt{\epsffile{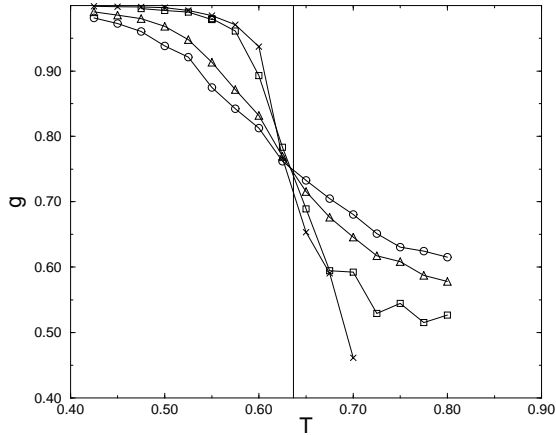}}
\end{center}
  \protect\caption[2]{Binder parameter $g$ as a function of $T$
for different sizes (in the low $T$ region, from bottom to top) 
$N=50,100,500,1000$. The curves become steeper
for larger values of $N$. The theoretical value of the bifurcation 
temperature $T=2/\pi$ is marked in the figure.
\protect\label{FIG2} }
\end{figure}

The order parameter can be calculated from Monte Carlo
simulations of the Hamiltonian (\ref{eq18}) with $\varepsilon(x)=|x|$ and 
$K_0=1$. The simulations use the heat-bath algorithm with a random 
sequential updating of the phases.
The transition temperature corresponding to 
the first branch is easily
obtained through standard finite-size scaling methods. Consider the
kurtosis (or Binder parameter) for the synchronization parameter $r$,
$g=\frac{1}{2}(3-\frac{<r^4>}{<r^2>^2})$,
where $<\ldots>$ is the standard configurational average [weighted with the
usual Boltzmann-Gibbs factor, $\exp(-\beta {\cal H})$, and ${\cal H}$ is given
by (\ref{eq18})]. The curves for $g$ are shown in figure 2 for different sizes.
Note that these curves (specially for $N=50,100,500$, data for $N=1000$ 
is more noisy) intersect at a common point characterizing the
bifurcation temperature.

Summarizing, we have considered models of oscillators interacting
through a general coupling function $f(x)$. In the absence of precessing
frequencies and for odd-coupling functions there exists a Lyapunov
functional. Then the probability density evolves toward stable
stationary states which can be described by an
equilibrium measure. In these cases the coupling force does not exert
work into the system and the dynamics is purely relaxational. 
We have
then proposed a family of exactly solvable models with singular
couplings.  The oscillators may synchronize more easily for less singular
couplings. In particular, case a), $f(x)=\delta'(x)$, never synchronizes,
case b), $f(x)=\delta(x)$, retains  synchronization only at zero temperature and
case c), $f(x)=$ sign$(x)$, syncronizes at a finite  temperature. This last case
is one of the few non trivial models for  synchronization dynamics which can be
analytically solved. Stationary  synchronized phases bifurcate from incoherence at a finite 
temperature which corresponds to a thermodynamic second order phase
transition. It would be very interesting to extend all our
considerations in the
presence  of $\omega_i$ for models with synchronization transition at finite
temperature where we expect the appearance of dynamical instabilities as
well as nonlinear behavior.

{\bf Acknowledgments}. This work has been supported by DGES (Spain) Projects
PB95-0296 (LLB) and PB95-1203 (JS), by FOM (The Netherlands) under contract
FOM-67596 (FR) and by European Union under TMR contract ERB FMBX-CT97-0157 (LLB
\& JS).

\hspace{-1.5cm}

\vfill

\end{document}